# Nanomechanical characterisation of a water-repelling terpolymer coating of cellulosic fibres

Julia Auernhammer[1$], Alena K. Bell[1$], Marcus Schulze[1], Yue Du[1], Lukas Stühn[1], Sonja Wendenburg[2], Isabelle Pause[3], Markus Biesalski[2], Wolfgang Ensinger[3], Robert W. Stark[1]*

*[1] Physics of Surfaces, Institute of Materials Science, Technische Universität Darmstadt, Alarich-Weiss-Str. 16, 64287 Darmstadt, Germany*

*[2] Makromolekulare Chemie und Papierchemie, Institute of Chemistry, Technische Universität Darmstadt, Alarich-Weiss-Str. 8, 64287 Darmstadt, Germany*

*[3] Materialanalytik, Institute of Materials Science, Technische Universität Darmstadt, Alarich-Weiss-Str. 2, 64287 Darmstadt, Germany*

*[$] Both authors contributed equally to this work.*

*Corresponding author:*

stark@pos.tu-darmstadt.de,

Phone: 061511621920

Fax: 061511621921

ORCID-IDs:
Julia Auernhammer   0000-0002-9896-7353
Alena K. Bell       0000-0001-7331-1879
Marcus Schulze      0000-0002-0009-3054
Lukas Stühn         0000-0002-7041-778X
Robert W. Stark     0000-0001-8678-8449

Abstract

Polymer coatings on cellulosic fibres are widely used to enhance the natural fibre properties by improving, for example, the hydrophobicity and wet strength. Here, we investigate the effects of a terpolymer P(S-co-MABP-co-PyMA) coating on cotton linters and eucalyptus fibres to improve the resistance of cellulose fibres against wetness. Coated and uncoated fibres were characterised by using scanning electron microscopy, contact angle measurements, Raman spectroscopy and atomic force microscopy with the objective of correlating macroscopic properties such as the hydrophobicity of the fleece with microscopic properties such as the coating distribution and local nanomechanics. The scanning electron and fluorescence microscopy results revealed the distribution of the coating on the paper fleeces and fibres. Contact angle measurements proved the hydrophobic character of the coated fleece, which was also confirmed by Raman spectroscopy measurements that investigated the water uptake in single fibres. The water uptake also induced a change in the local mechanical properties, as measured by atomic force microscopy. These results verify the basic functionality of the hydrophobic coating on fibres and paper fleeces but call into question the homogeneity of the coating.

Acknowledgement

The authors would like to thank the Deutsche Forschungsgemeinschaft under grant PAK962 project-id 405549611 and 405422473 for the financial support.

**Declaration of the conflict of interest**
There are no conflicts of interest to declare.

# 1 Introduction

Introducing paper as a functional material in the fields of microfluidics, electronics, sensor technologies, and medicine is equally promising and challenging (Bump et al. 2015; Delaney et al. 2011; Gurnagul and Page 1989; Hayes and Feenstra 2003; Liana et al. 2012; Ruettiger et al. 2016). In particular, material swelling and the loss of mechanical stability in a humid or wet environment must be addressed. How the relative humidity (RH) affects the elastic modulus, stiffness and strength of materials has been studied by various authors (Ganser et al. 2015; Placet et al. 2012; Salmen and Back 1980). To control the wetting properties, paper fibres or the entire fleece can become hydrophobic with the application of polymer coatings (Janko et al. 2015; Jocher et al. 2015) or modified with $TiO_2$ nanoparticles or fluorinated silanes to obtain self-cleaning surfaces used for the separation of different liquids, such as water and oil (Chauhan et al. 2019; Satapathy et al. 2017; Tudu et al. 2019; Tudu et al. 2020). To improve the strategies to strengthen wetting properties, it is essential to understand how these coatings change the mechanical and wetting properties of individual cellulosic fibres.

The efficiency of coating strategies and the uptake of liquids into cellulose fibres are issues that have been successfully addressed with various characterisation methods. The distribution of coatings on fibres and the distribution in the fibre network can be investigated with fluorescence microscopy (Bump et al. 2015; Janko et al. 2015). Raman spectroscopy was applied by (Fechner et al. 2005) and (Eronen et al. 2009) to study changes in the molecular structure of cellulose caused by water or sodium hydroxide. In a more comprehensive approach, coated cellulose fibres were investigated with the help of brightfield microscopy, Raman spectroscopy and confocal laser scanning microscopy (Janko et al. 2015). Atomic force microscopy (AFM) has been used to investigate the surface properties of cellulose-based materials in dry and wet states (Chhabra et al. 2005; Li et al. 2020). Additionally, the viscoelastic properties of pulp fibres could be investigated with AFM. Here, the RH was varied, and the elastic moduli and viscosity characteristics of fibres were examined. Compared with the dry state, the elastic moduli decreased by a factor of 10, and the viscosity decreased by a factor of 10-20. When immersed in water, the fibres exhibited a decrease in the elastic moduli by 100 and in the viscosity by at least three orders of magnitude (Czibula et al. 2019). Further AFM-based colloidal probe measurements on cellulose were performed on gel beads made of cellulose to show the impact on the mechanical properties in the wet state (Hellwig et al. 2017). With an AFM-based indentation method, it was possible to test the mechanical properties of wet cellulose and observe the Young's modulus, which resulted in elucidating the kPa range (Hellwig et al. 2018). Further nanoindentation experiments have been carried out to study the elastic modulus of wood cells (Gindl and Schoberl 2004) or that of dry pulp fibre walls (Adusumalli et al. 2010). In 2013, Yan *et al*. used the classic Oliver-Pharr approach (Oliver and Pharr 1992) for the investigation of the instantaneous elastic modulus (Yan and Li 2013). In 2014, *Ganser et al*. described the hardness and modulus of pulp fibres via AFM-based nanoindentation as a function of the relative humidity (Ganser et al. 2014) and, in 2015, compared the results with viscose fibres (Ganser et al. 2015). Additionally, the breaking load of a single fibre as a function of the RH could be determined, and the breaking load decreased with increasing RH (Jajcinovic et al. 2018).

Herein, we report on the effect of humidity on unmodified cotton linter paper (LP) and eucalyptus sulphate paper (EP) fibres, as well as LP and EP fibres coated with a fluorescent and hydrophobic polymer, namely, (polystyrene-co-4-methacryloxybenzophenon-co-1-pyrenemethylmethacrylate) P(S-co-MABP-co-PyMA), which is a promising terpolymer to improve the hydrophobicity of paper fleeces and fibres. Additionally, its fluorescent component allows for the mapping of the polymer distribution. For the following experiments, papers made of LP and EP fibres were coated, and their properties in comparison to those of unmodified paper samples were investigated. LP consists of 94-95 % cotton and serves as a model system (Mather and Wardman 2015; Young and Rowell 1986). The chemical composition of EP is more complex than that of LP, with EP having a cellulose content of only 42 % and additional hemicellulose and lignin (Young and Rowell 1986). The functionality of the water-repelling polymer coating was investigated by macroscopic (contact angle (CA) goniometry and fluorescence microscopy) and microscopic (Raman spectroscopy, scanning electron microscopy (SEM) and AFM) methods to reveal the response of the different samples to humidity at a large scale.

## 2 Materials and Methods

### 2.1 Materials

Cellulose fibres were extracted from LP and EP sheets, which were prepared according to DIN 54358 and ISO 5269/2 (the Rapid-Köthen process). The hydrophobic, UV-active, and fluorescent terpolymer P(S-co-MABP-co-PyMA) was used as the coating. Specifically, the PyMA part of the terpolymer was fluorescent. The coating was applied by dip coating the paper samples, according to the procedure described in the literature (Boehm 2014; Böhm et al. 2013). Briefly, the polymer was dissolved in tetrahydrofuran (THF) and in a solution applied *via* the dip coating process. After drying, the coated paper sheets were irradiated by UV light (365 nm), and excess polymer was removed by THF Soxhlet extraction; this suggests that all remaining fibre-attached polymers are chemically stable, and we propose that it is thermally stable up to 300 °C, as the polymer is an organic material (Boehm 2014; Böhm et al. 2013; Wendenburg et al. 2017).

### 2.2 Scanning Electron Microscopy

A conducting gold layer was sputtered (Balzer SCD 050, BAL-TEC AG, Balzers, Lichtenstein) onto the samples with a supply current of 30 mA for 75 s. For SEM imaging, a Philips XL30 FEG high-resolution electron microscope was used with an electron voltage of 10 kV at 1500x magnification and 10000x magnification (figure S1).

### 2.3 Fluorescence Microscopy

Fluorescent images were taken using a Zeiss AXIO observer Z1 optical microscope equipped with a Zeiss Colibri.2 illumination system and a Zeiss HSM Axiocam camera (Carl Zeiss Microscopy GmbH, Jena, Germany). All images were processed using Zeiss Axiovision SE64 software. Images were taken in the standard epifluorescence configuration and in transmitted white light mode for bright field microscopy. The excitation wavelength was 365 nm, and the emission bandwidth was 435-485 nm (AHF Analysetechnik, Tübingen, Germany, filter set 31-013, DAPI configuration). The acquisition time for the fluorescent light was 1 s with a 5x/0.12 objective lens for the unmodified samples and 10 s with a 5x/0.12 objective lens as well as 2 s with a 10x/0.25 objective lens for the coated samples.

Pieces of paper and small bundles of fibres were analysed by pulling out fibre bundles of the pieces of paper. Papers and fibres were placed on a microscope slide and covered with a cover glass.

## 2.4 Contact Angle Goniometry

For the contact angle goniometry (CA) measurements, a CA goniometer (OCA, DataPhysics Instruments, Filderstadt, Germany) was used. A static Milli-Q water droplet with a volume of 2 $\mu$l was placed on the sample. The CA was determined *via* Young-Laplace alignment 1 s after drop generation.

## 2.5 Raman Spectroscopy

A confocal Raman microscope (alpha300R by Witec GmbH, Ulm, Germany) with a laser wavelength of 532 nm (Nd:YAG-Laser) and a 20x/0.40 objective lens was used throughout the measurements. For reference measurements, small pieces of paper were collocated on an object slide to obtain unmodified samples. Raman measurements were taken with 10 accumulations and an integration time of 30 s each.

Point scans (10 accumulations with an integration time of 10 s each) and map scans (25 x 25 $\mu m^2$, 50 x 50 measuring spots with an integration time of 0.005 s each) were acquired for varied states of humidity (humidity experiments). The measurements were taken inside the fibre close to the surface to estimate the amount of water intruding into the fibre.

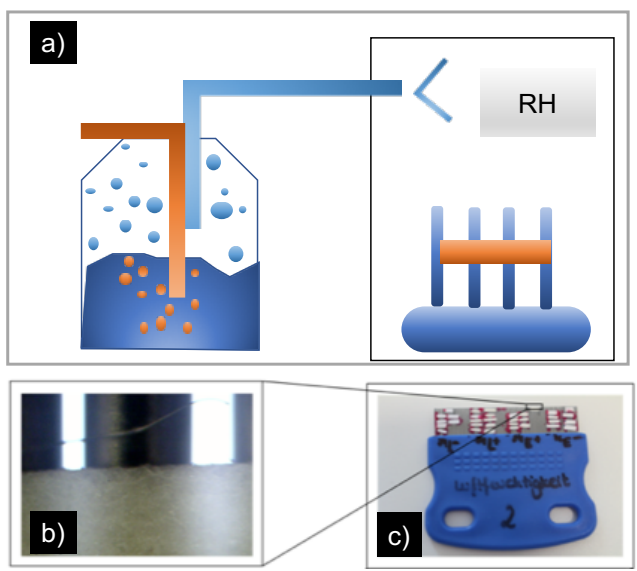

**Figure 1** (a) Experimental setup for humidity experiments including a water wash bottle, a N₂ gas inlet, a hygrometer (RH = relative humidity), and a small tooth comb as the sample holder. The first inset (b) is a magnification (5x objective) of two teeth of the comb with the paper sample, and (c) shows a photographic image of the sample holder.

For the humidity experiments, the paper sheets (2 x 7 mm$^2$) were mounted on a small-tooth comb that was placed in a climate chamber (figure 1). For every type of sample, three sets of experiments were performed, in which the spectral range from 3200-3500 cm$^{-1}$, corresponding to the OH stretching vibration, was analysed. In this way, the water uptake in the LP, polymer-coated LP (PCLP), EP, and polymer-coated EP (PCEP) samples could be recorded. The calculation of the intensity in this region was performed by integration (integrated intensity).

In the first set of experiments, point scans were used, and the Raman spectra were measured first under room conditions (20-60 % RH at 22-24 °C). The second point scan was performed after keeping the sample at 80-90 % RH for 60 min. Subsequently, the RH was reduced to 2-5 %, and the third point scan was recorded after another 60 min.

The second set of experiments included map scans every 60 min. To evaluate potential local drying caused by the heat of the Raman laser, map scans were also performed because the laser radiation is distributed over a large area. The first measurement was taken under room conditions, and the second measurement was taken after 60 min at 80-90 % RH. For the third map scan, a drop of water was added to the sample, and after 15 min, a Raman spectrum was acquired. The last measurement was taken after 60 min at 2-5 % RH.

In the third set of experiments, the adjustment time for RH was 180 min so that the first measurement was taken under room conditions, the second after 180 min at 80-90 % RH, and the last after 180 min at 2-5 % RH.

All experimental data were processed by using WITec Project FOUR 4.0 software. The background of the point scan spectra was subtracted, and the spectra were normalised to the Raman band at 2800-3000 cm$^{-1}$. From the map scans, an average spectrum of all measurements was calculated and filtered for illustration. From this region, the background was subtracted by a 2$^{nd}$-order polynomial. From all map scans, an average spectrum was calculated that was then treated like a point spectrum. In this way, the map scans could be compared directly to the point scans.

For the error calculation, the following equation was used:

$$\Delta I = \frac{I}{A} \cdot \Delta A \qquad (1)$$

where I is the intensity and A is the Raman band height (amplitude).

## 2.6 Atomic Force Microscopy

### 2.6.1 Indentation Experiments

A Dimension ICON instrument (Bruker, Santa Barbara, CA, USA) was used for the indentation and nanomechanical mapping experiments on freely suspended fibres. The indentation experiments were performed using ZEIHR cantilevers with a force constant of 27 ± 4 N/m. The set point was chosen as 300 nN, and the engagement rate was 0.5 Hz. The indentation was recorded *via* static deflection versus separation curves. The inverse optical lever sensitivity (InvOLS) was extracted from every curve, as shown in figure 2b. The InvOLS is a relative measure of fibre stiffness plus the cantilever stiffness, with units of nm/V. Thus, the recorded deflection versus separation curves were representative of the resistance of the fibres towards indentation by the AFM probe. Variations in the InvOLS linearly correspond to changes in the sample stiffness. Dividing the recorded InvOLS by the calibrated InvOLS of the cantilever and its force constant allows for the representation of the data in nm/nN, which shows the indentation per applied force (see equation 2). The cellulose fibres were manually extracted from paper sheets and adhered onto a sample holder that provided a gap between the two attachment points of the fibre. To straighten the fibres, the distance of the gap was adjusted by a micrometre screw. Deflection versus separation curves were acquired at L/2 of the freely suspended fibre to minimise any influence of the mounting and prevent any effects due to an asymmetric load. For every type of sample, two results are presented, and every data point is the average from three deflection-versus-separation curves. Two measurement points were taken under room conditions (region I). After hydration of the fibre (a drop of deionised water, which was carefully removed after 30 min of swelling time), data points were acquired in 10-min intervals for a period of 200 min (region II). Then, the fibre was dried in a gentle stream of nitrogen, and two additional measurements were performed (region III). The fibres were freely suspended to homogeneously expose the fibre to humidity/dry air and to guarantee the measurement of the mechanical properties of a single fibre without the influence of a substrate or other connecting fibre bonds.

$$\text{Indentation}\left(\frac{\text{nm}}{\text{nN}}\right) = \frac{\text{InvOLS (Curves)}\left(\frac{\text{nm}}{\text{V}}\right)}{\text{InvOLS(Cantilever)}\left(\frac{\text{nm}}{\text{V}}\right)\cdot k\left(\frac{\text{N}}{\text{m}}\right)} \qquad (2)$$

### 2.6.1 Nanomechanical mapping

For nanomechanical mapping, PeakForce-Tapping mode was applied using a ScanAsyst-Fluid+ cantilever from Bruker (Santa Barbara, CA, USA) with a spring constant of 0.9 ± 0.2 N/m. The setpoint was chosen as 3 nN, the scan rate was 0.5 Hz, and the amplitude was 300 nm. Single paper fibres were manually extracted from paper sheets, adhered onto 3D-printed fibre holders and put inside a climate chamber of the AFM setup. When measuring the polymer-coated fibres, the paper fibres were primarily investigated with fluorescence microscopy to detect the homogeneous polymer-coated spots on the fibre surface. The chosen RHs were 2 %, 40 %, 75 % and wet conditions (a drop of deionised water on the fibre). The RHs were maintained for 30 min. The deionised water drop was applied and left for 30 min and was subsequently removed carefully with clean tissue. As shown in (Carstens et al. 2017; Hubbe et al. 2013) the water progress into cotton linters test stripes goes in seconds. Furthermore, (Olejnik 2012) discovered that the free swelling time for a whole pulp was 70 minutes. However, (Mantanis et al. 1995) recorded the swelling of cellulose single fibres in water and showed that the equilibrium is reached at 45 minutes. As we are interested in the fibre surface only, we suggest that 30 minutes as "swelling time" for every hydration condition is sufficient. Additionally, the water drop for swelling was removed before the measurement, where the fibre could not swell further then.

The fibre and AFM head were not immersed in water during the measurements. The topography images taken were tilt- and drift-corrected by a first-order plain fit. The mechanical property images were not modified. For analysis of the AFM images, NanoScope Analysis software from Bruker was used. For each fibre type, three different individual fibres were investigated, and on each fibre, three spots were mapped and analysed.

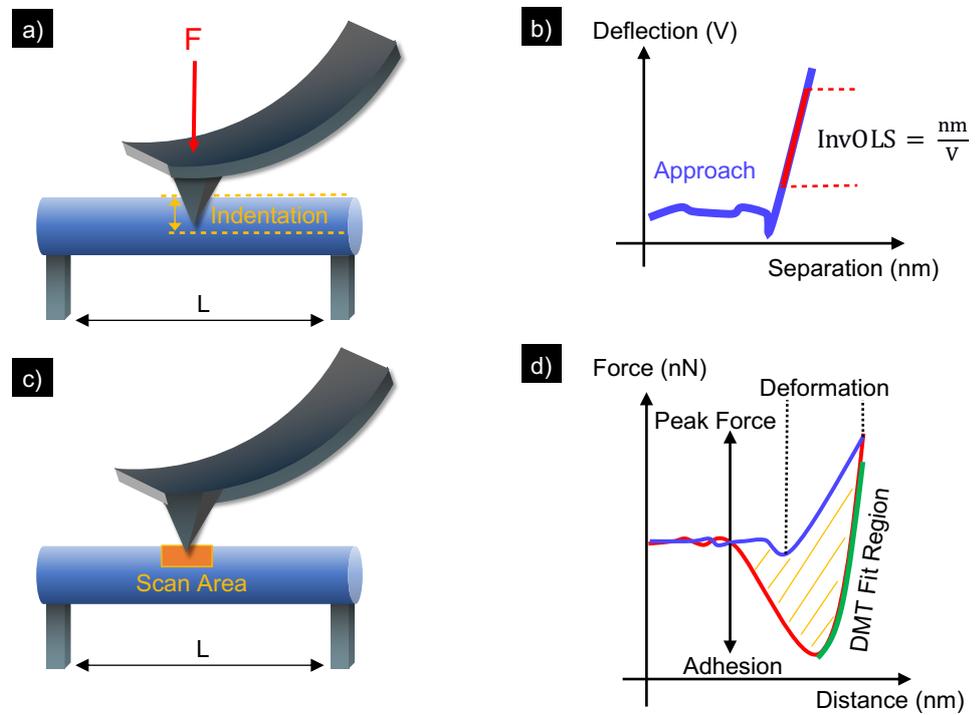

**Figure 2** Schematic setup of AFM experiments. (a) Schematics of nanoindentation experiments, (b) deflection versus separation curve with InvOLS shown, (c) schematics of nanomechanical mapping and (d) force versus distance curve with extracted mechanical properties.

# 3 Results and Discussion

## 3.1 Scanning Electron Microscopy

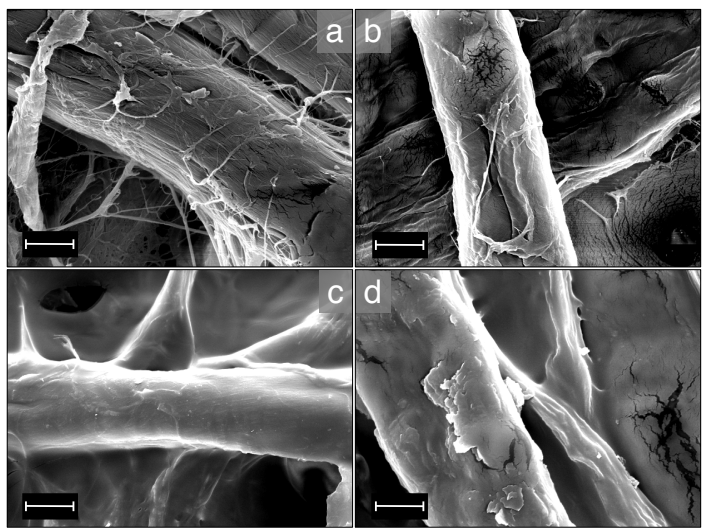

**Figure 3** SEM images of sheets made of (a) LP, (b) (EP), (c) PCLP, and (d) PCEP. (scale bar: (a,b,d) 5 µm and (c) 10 µm)

SEM was used to visualise the structure of LP and EP and their morphological changes induced by the polymer coating. For all samples, the paper fibres were homogeneously distributed (figure S1). LP exhibited an increased level of entanglement (figure 3a) due to a higher number of smaller fibrils than EP (figure 3b). The polymer coating wrapped the fibres, which was very obvious for the PCLP (figure 3c), thereby changing the appearance of the small entanglements from single fibres to agglomerated structures. For EP, the number of entanglements also decreased (figure 3d) because fibres with a smaller diameter were adhered onto fibres with a larger diameter. In figure S1, overview SEM images at a 1400x magnification are given of all paper and paper-polymer systems.

## 3.2 Contact Angle Goniometry

With the help of static CA measurements, the hydrophobic character of the polymer coating was investigated on a macromolecular scale. The unmodified paper samples adsorbed the water quickly, and videos were recorded to obtain the measurements. These videos were stopped 1 s after the water drop was applied to determine the CA of the different samples. The images of sessile droplets (2 $\mu$l of deionised water) on the LP, PCLP, EP, and PCEP sheets for the CA measurements are given in figure 4.

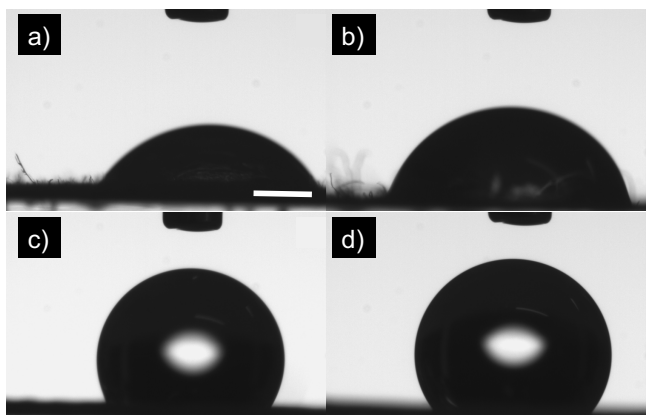

**Figure 4** Images of 2 µl water drops on (a) LP, (b) EP, (c) PCLP, and (d) PCEP for CA measurements. The scale bar is 500 µm.

LP and EP showed a hydrophilic behaviour. The imbibition was faster in EP (2 s) than in LP (8 s). On both unmodified samples, the CAs were in the same regime, which was also the case for the polymer-coated samples. The CAs of PCLP and PCEP were in the hydrophobic range, with value of 119° ±10° and 129° ± 9°, respectively, while the CAs of LP and EP showed hydrophilic behaviour with CA values of 79° ± 10° on both materials. On the polymer-coated samples, the water drop was not adsorbed and was much more stable than in the unmodified samples. Additionally, figure 3 shows a smoother surface for the polymer-coated samples than for the unmodified samples, which was also observed in the SEM measurements (figure 3). These results proved the hydrophobic character of the polymer coating.

### 3.3 Fluorescence microscopy

The distribution of the polymer on the paper was analysed by fluorescence imaging. The measurements were taken on paper fleeces as well as on small fibre bundles to evaluate the distribution of the polymer. In figure 5, overlays of bright field microscopy images (grey) and the corresponding fluorescence microscopy images (yellow) are shown.

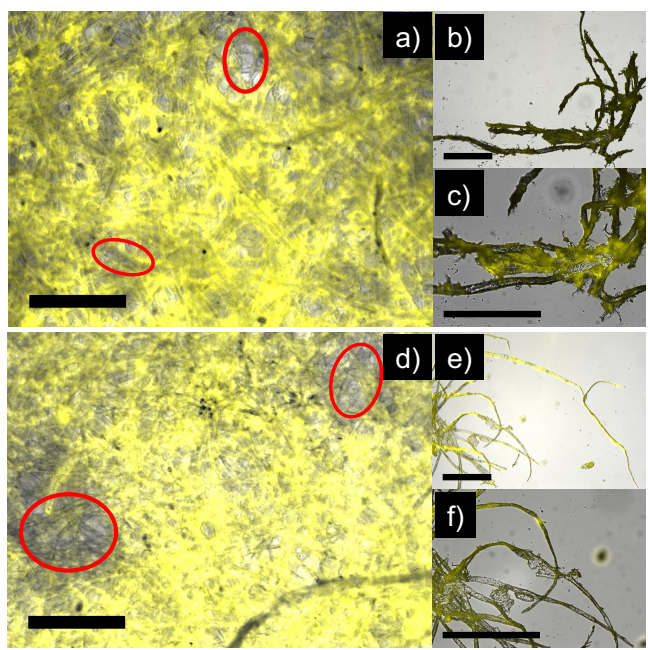

**Figure 5** Overlay of bright field microscopy (grey) and fluorescence microscopy images (yellow) at 5x and 10x, showing sheets of (a) PCLP and (d) PCEP, as well as near-surface fibres of (b,c) PCLP and (e,f) PCEP. The scale bar represents 300 μm. The red areas show regions without a polymer coating, therefore revealing the inhomogeneous coating distribution.

A faint and evenly distributed fluorescence could be seen for the unmodified samples, which was caused by the autofluorescence of the lignin (figure S2). A higher autofluorescence was noticed on EP because of the higher lignin content in EP than in LP (Young and Rowell 1986). In contrast, PCLP and PCEP exhibited a strong fluorescent signal that was localised along the fibres. Nevertheless, an analysis of the fluorescent signal showed that the fleece was not homogeneously coated (red areas in figure 5 a,d). Additionally, on the level of single fibres, as shown in figure 5c,f, the fluorescence signal and thus the amount of coating could vary. The inhomogeneous polymer distribution implies that water could infiltrate not only the unmodified paper but also the coated samples.

### 3.4 Raman Spectroscopy

Raman measurements were performed to investigate the water uptake in the samples as a function of the moisture and incubation time by observation of the OH band. The reference Raman scans, including the normalised intensity versus Raman shift graphs with the respective vibrational motions, are given in figure 6. A table of the peak assignments can be found in Table S1. Briefly, the vibrational motions of carbon and oxygen are located in the lower wavenumber regime (379-1118 cm$^{-1}$), while peaks at higher wavenumbers (1290-3500 cm$^{-1}$) are due to vibrations involving carbon, hydrogen, and oxygen.

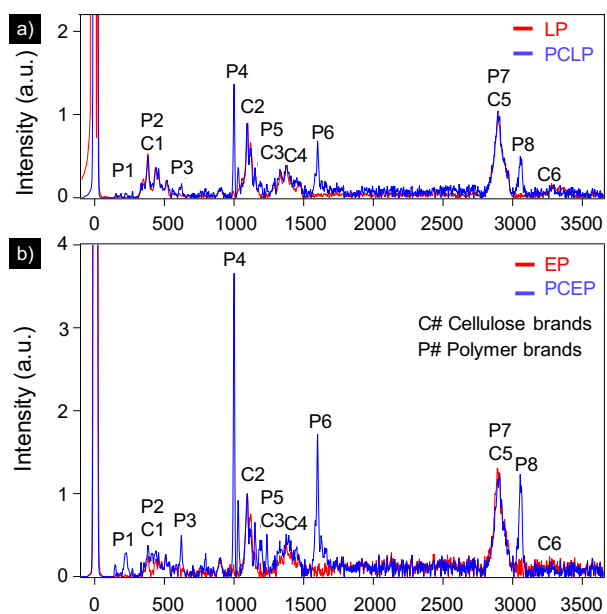

**Figure 6** Raman spectra of (a) LP and PCLP and (b) EP and PCEP with assigned cellulose (C#) and polymer (P#) Raman bands. Unmodified paper is given in red, and polymer-coated paper is given in blue. The spectra were taken at 20x/0.4 with 10 accumulations and a 10-s integration time each. The measurements were taken on the paper surface where there was a crosslink of paper fibres. The peak assignments can be found in Table S1.

Raman spectra of polymer-coated samples (figure 6, blue spectra) showed the superposed spectra of paper (figure 6, red spectra, C# labels) and the polymer (figure 6, P# labels). The Raman spectrum of the polymer blend P(S-co-MABP-co-PyMA) (figure S3) shows the superposition of the components PS, benzophenone (BP), and pyrene. For the peak assignments, the peak positions were compared to Osterberg *et al.* for cellulose (Osterberg et al. 2006), Brun *et al.* and Hong *et al.* for PS (Brun et al. 2013; Hong et al. 1991), Babkov *et al.* for BP (Babkov et al. 2006), and Xie *et al.* for the pyrene component (Xie et al. 2010). The peak positions for all materials are also listed in Table S1.

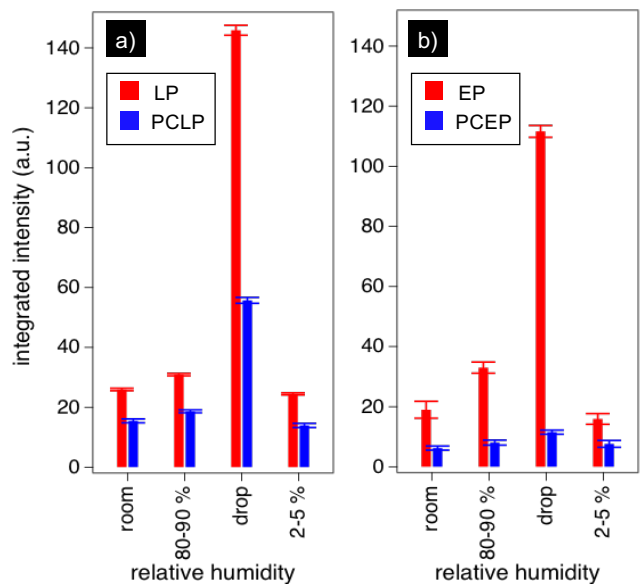

**Figure 7** Relative peak intensities of the band assigned to OH stretching obtained in hydration experiments with in situ Raman spectroscopy for (a) LP and PCLP and (b) EP and PCEP. Data of the unmodified paper samples are represented in red (left bar), and the polymer-coated samples are represented in blue (right bar). The integrated intensity is the summed intensity of OH stretching in the spectral region of 3200-3500 cm$^{-1}$.

To characterise the water uptake, the integrated intensity of the OH band was monitored. Specifically, the OH band was monitored in a region of interest inside the paper fleece but close to the surface. The integrated intensity versus RH graphs in figure 7 and figure S4 correlate with the water uptake. In figure 7, the integrated intensity is given for room conditions (50 % RH), 80-90 % RH, conditions in which the sample was fully wetted by a droplet and dry conditions at 2-5 % RH. For the unmodified samples, the integrated intensity of the OH peak was substantially higher after adding the drop of water in comparison to measurements at 80-90 % RH for 60 min or 180 min. This led to the result that the time of exposure to 80-90 % RH was not long enough to fully wet the paper samples. Comparing the experiments after adding the water drop to unmodified samples, it was observed that the water uptake of EP was higher than that of LP, which agreed with the CA goniometry results. The more pronounced uptake of water for EP was explained by the high content of hemicellulose in comparison to LP (Young and Rowell 1986).

For the modified samples, the change in the integrated intensity was predicted to be less distinct than in the unmodified samples because the water should not enter the paper through the polymer coating. In fact, the hydrophobicity of the coating could be confirmed by observing the sliding motion of drops over the sample surface, therefore not being imbibed. Thus, the addition of the water drop did not lead to as much of an increase in the OH peak intensity as that for the unmodified samples. For an increased RH, the change in integrated intensity was also smaller than for the unmodified samples.

The change in integrated intensity was larger for the unmodified samples than for the polymer-coated samples. The results supported the assumption of a hydrophobic character of the cellulose samples after the addition of the polymer coating.

## 3.5 Atomic Force Microscopy

### *3.5.1 Indentation Experiments*

To characterise the variation in the mechanical characteristics of the fibres in the different states of hydration, AFM indentation measurements were performed on individual fibres.

The indentation versus time diagrams were divided into three regions, according to three different states of hydration: (I) fibres under room conditions, (II) soaked fibres, and (III) nitrogen-dried fibres. A higher indentation depth at the fixed deflection point was correlated with fibre softening. The depicted changes in percentage were extracted from the average values of the respective region. The results are shown in figure 8a,b. EP fibres underwent softening (76 ± 13 %) from region I to region II with a consequent continuous increase in the indentation. In contrast, PCEP fibres not only showed an increased stiffness in region I but also a reduced softening in the hydrated state (12 ± 6 %) in comparison to EP. The direct comparison of EP and PCEP fibres showed that the polymer coating increased the stiffness of the fibre. The reduced softening was related to reduced water uptake. The indentation values for both samples returned to their respective values (region I) after dehydration (region III). Softening of the LP fibre (38 ± 15 %) when hydrated could be measured. The PCLP fibres were in the same range as the PCEP fibres (0.1 – 0.3 nm/nN) and showed the same softening behaviour (12 ± 5 %) upon hydration. LP fibres showed the highest stiffness among the samples, which was attributed to the higher content of cellulose contained in LP than in EP. This agrees with Baley *et al.* and Gassan *et al.*, who stated that the elastic modulus of natural fibres increases with increasing cellulose (Baley 2002; Gassan et al. 2001; John and Thomas 2008). Another possible reason for this result is the high portion of crystalline regions in linters, while eucalyptus mostly has amorphous sites. The crystalline regions are expected to be stiffer than amorphous regions, which could explain the high resistance of the LP fibres against indentation compared to that of the EP fibres.

It can be concluded that the coated samples show less softening when exposed to water than the uncoated samples, which is correlated with less water uptake. Furthermore, the coated samples exhibited a very similar behaviour independent of the underlying fibre material. While the indentation experiments at a single point confirm the functionality of the polymer coating as a water barrier, further AFM experiments are necessary.

In comparison to earlier studies of AFM nanoindentation on fibres (*e.g.,* Ganser *et al.* 2015), the fibres presented here were freely suspended, and their surface was not supported. This setup introduced further degrees of freedom for the fibres to react towards a load, including bending and twisting. Although deflection versus separation curves that indicated movements of the fibre were not considered for evaluation, the combination of probe indentation with movement of the fibre could not be fully excluded. Thus, nanoindentation experiments were not used to calculate the reduced modulus but to show a force-dependent displacement as a measure of mechanical resistance. The results allowed for a qualitative rather than a quantitative comparison between the different states of hydration and thus supported the presented results of the further applied characterisation techniques.

A more comprehensive understanding of the samples requires two-dimensional maps that show the topography and its correlation with the mechanical properties. An identification of crystalline and amorphous regions and their respective modulus as a function of various states of hydration could support or refute the explanation of a stiffer LP due to crystallinity. Additionally, the acquired distribution of the modulus provided by two-dimensional maps can be used to evaluate the representativeness of the previously presented values. Thus, PeakForce-Tapping mode was applied to freely suspended fibres.

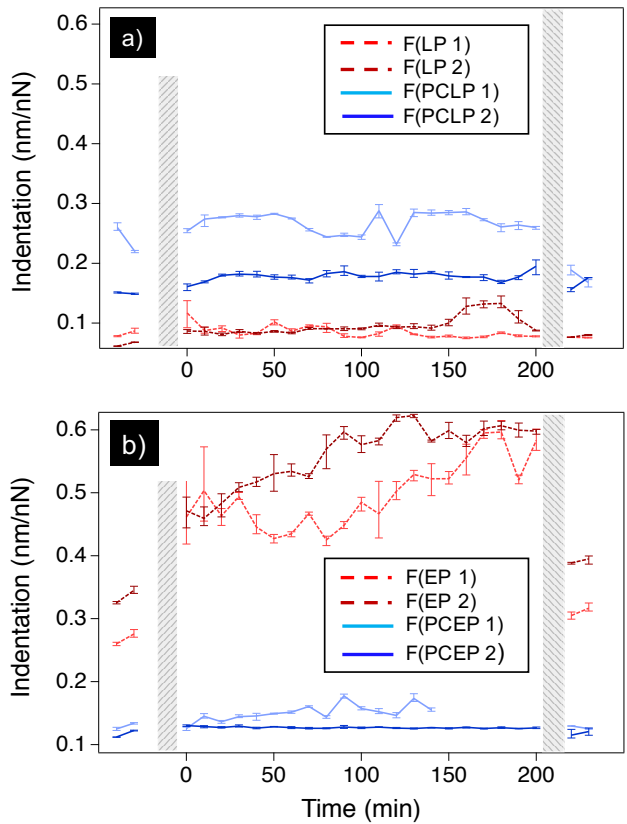

**Figure 8** Mechanical resistance against indentation of the AFM probe into (a) LP and PCLP and (b) EP and PCEP in the setup with a fixed distance between attachment points (F). Region I depicts the state under room conditions, region II shows the soaked state and region III is the after-drying state.

### 3.5.2 Nanomechanical Mapping

To investigate the mechanical properties of the paper fibres as a function of the RH, AFM measurements were performed on freely suspended single fibres. For the polymer-coated fibres, fluorescence microscopy was applied before the AFM measurements to identify homogenously coated regions of interest on the paper fibre.

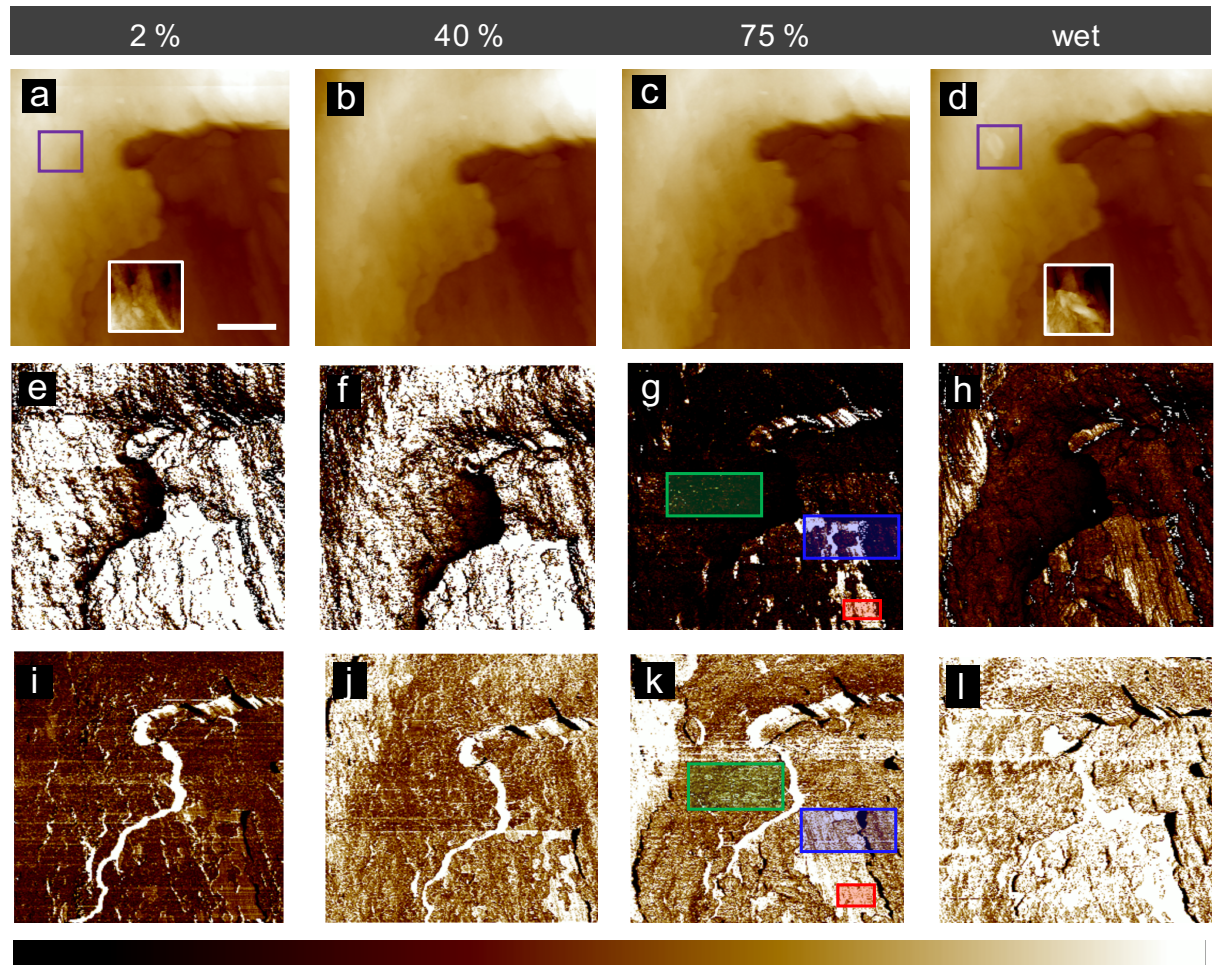

**Figure 9** AFM images of LP fibres. The RH was varied from 2 % to a wet condition. (a-d) Topography images, (e-h) DMT modulus maps and (i-l) adhesion maps. The scale bar is 800 nm. The colour bar is -1 – 1.3 µm for topography images, 0 – 3 GPa for DMT modulus maps and 0 – 2 nN for adhesion maps. In the white framed topography images, the colour scale was modified to 0 – 250 nm. The white boxed images in a) and d) show the topography of the swollen microfibrils. An amorphous region of cellulose is highlighted in green, while the crystalline region is marked in red, and a transition between crystalline and amorphous regions is shown in blue.

The quasi-static PeakForce-Tapping mode allows for the simultaneous mapping of the topography and mechanical properties. Here, cumulative force versus distance curves were measured at every pixel in the recorded map. The mechanical properties were then directly extracted from the force versus distance curves. While the adhesion force represents the minimum in the retraced curve, the DMT modulus was fitted, as shown in figure 2b, with the prediction of the contact mechanics by Derjaguin, Muller, and Toporov (DMT) (Derjaguin et al. 1975). In contrast to other contact mechanics models, such as the Johnson, Kendall and Roberts (JKR) model (Johnson et al. 1971), the DMT model includes adhesion outside the contact area and is suitable for high elastic moduli, low adhesion and a small radius of indentation. Therefore, the DMT model was used for the contact mechanics in the measurements. As shown in figure 9, the LP fibre exhibited crystalline and amorphous regions in the topography image, but especially in the mechanical property maps, such as the DMT modulus and adhesion, the differences in the crystal structure of cellulose were clearly distinguishable. In the mechanical property maps, the adhesion increased with increasing RH, while the DMT modulus decreased. Overall, swelling of microfibrils was observed in the AFM topography images as the RH was increased. Figure 9a shows the topography image with the microfibrils at 2 % RH. Comparing the microfibrils in figure 9a to those in figure 9d, which shows a fully hydrated fibre, the microfibrils appear to be swollen under wet conditions. The white box (figure 9a, d) indicates in detail the swelling of microfibrils due to absorbed water inside the cellulose network. The insolubility of cellulose results in swelling of the microfibrils (Lindman et al. 2010). Water infiltrates the network and diffuses through the pore space system by capillary condensation. Therefore, the volume increases, which leads to an expansion in the crystal structure and eventually breaks the hydrogen bonds in the cellulose network (Cabrera et al. 2011; Gumuskaya et al. 2003; John and Thomas 2008). Thus, variations in the RH can induce changes in the shape of a microfibril due to the not entirely connected cellulose molecules, as shown in figure 9d, in the marked areas. Crystalline and amorphous regions showed different values for the DMT modulus, which is in agreement with Cabrera *et al*. (Cabrera et al. 2011). Cross sections of the DMT modulus versus the position over a crystalline and an amorphous region from figure 9 are shown in figure 10b,c. The crystalline region showed a DMT modulus of 80-100 GPa, while the amorphous region exhibited a DMT modulus of 20-30 GPa. The transition from crystalline to amorphous structures, and thus the transition in the mechanical properties, seemed to be sharp and well defined according to figure 10a.

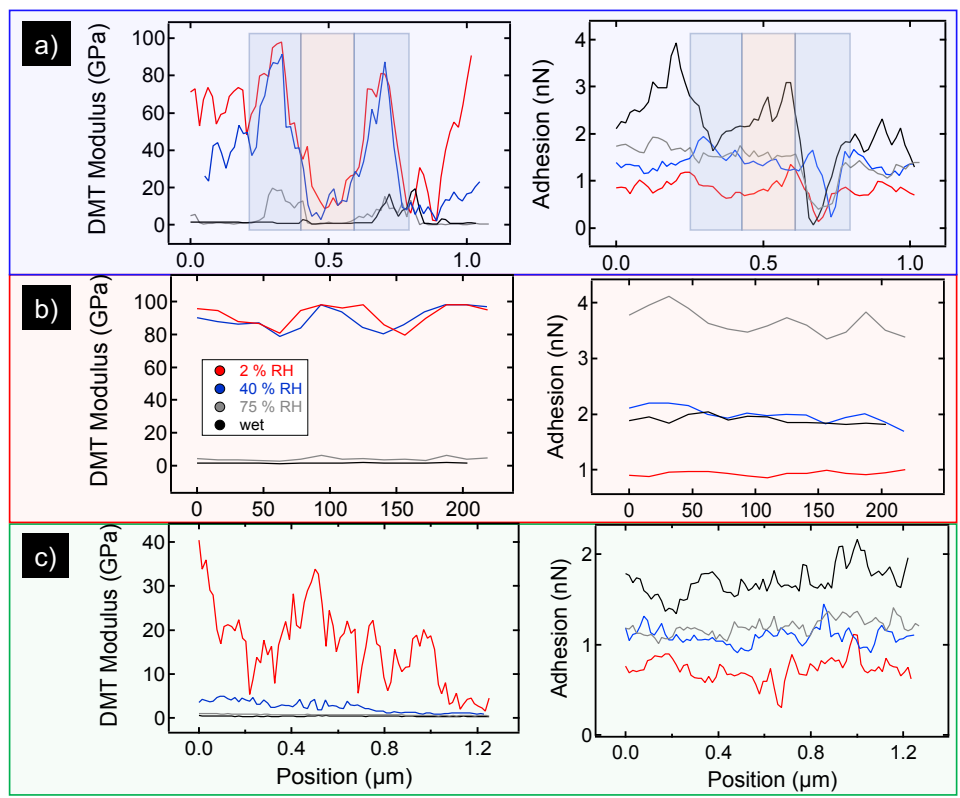

**Figure 10** Cross sections corresponding to figure 8 of the DMT modulus in (a) a transition area between crystalline, amorphous and crystalline regions. The crystalline regions are highlighted in light blue, and the amorphous region is highlighted in light orange. The cross sections of the DMT modulus and adhesion of a crystalline region are represented in (b), whereas (c) illustrates this for an amorphous region.

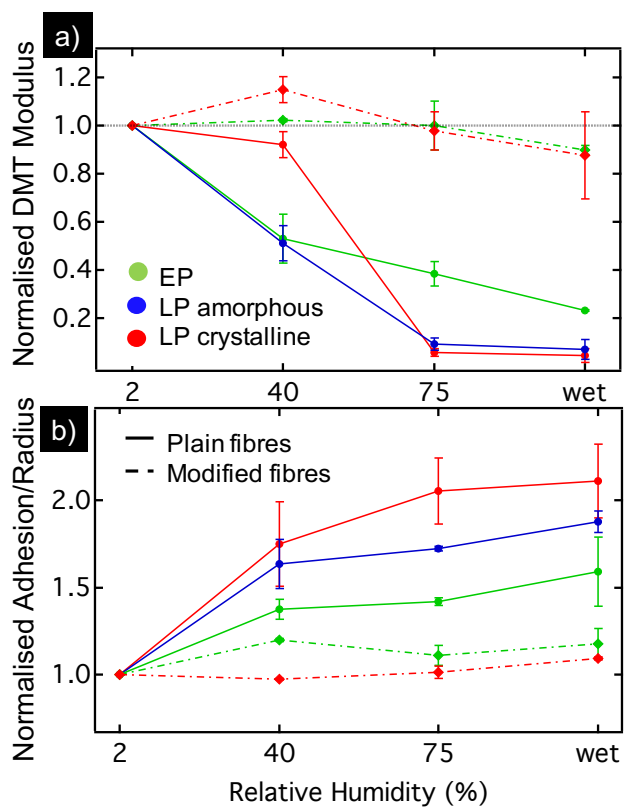

**Figure 11** a) Normalised DMT modulus and b) normalised adhesion/radius plotted against the relative humidity in unmodified fibres and polymer-coated fibres

A general decrease in the normalised DMT modulus with increasing RH was observed in crystalline and amorphous LP fibres, as well as in EP fibres (figure 11a). Up to 40 % RH, the crystalline regions of the LP fibres did not show a significant decrease in the normalised DMT modulus, as recorded for the amorphous regions. If the RH was sufficiently high, softening of the crystalline regions was also initiated. More water molecules intrude the fibre network, diffuse to the crystalline areas and break the hydrogen bonds in these ordered areas (Persson et al. 2013). If the fibre reached a fully hydrated state, both the crystalline and amorphous regions lost their stability. A similar increase in the normalised adhesion/radius in crystalline and amorphous regions was observed, as shown in figure 11b. When increasing the RH to wet conditions, the crystalline regions had a higher value in the normalised adhesion/radius than the amorphous regions. However, this deviation lies between the error bars and is therefore not interpreted as being distinct. With the increasing amount of water molecules inside the hydrophilic LP fibre, attractive capillary forces accumulate between the AFM tip and the LP fibre surface. Because of the softening the fibre, i.e., the decreasing DMT modulus, the tip was indented deeper into the surface, which led to the requirement of a higher reset force of the tip and therefore to a higher measured adhesion. The EP fibres consist of a more complex system than the LP fibres. While LP fibres possess a cellulose percentage of 95 %, EP fibres also have portions of hemicellulose and lignin in their system (Mather and Wardman 2015; Young and Rowell 1986). Cellulose is embedded in an amorphous matrix of hemicellulose and lignin. Hemicellulose is bonded *via* hydrogen bonds to cellulose and acts as a cementing matrix. Additionally, hemicellulose is not crystalline and has a hydrophilic character. Lignin acts in EP fibres as a hydrophobic network, which increases the stiffness and is totally amorphous. Figure 11a shows a decrease in EP fibres in the normalised DMT modulus when increasing the RH. At 40 % RH, the percentage of decrease in the normalised DMT modulus is similar to that of amorphous cellulose. Since the EP fibre mostly consists of noncrystalline parts, the fibre softens in large parts due to infusing water molecules breaking the hydrogen bonds between cellulose molecules. In contrast to simple amorphous cellulose, the EP fibres showed a linear decrease in the normalised DMT modulus when the RH was further increased to a wet condition, as indicated in figure 11a by the green line. This linear decrease was assumed to be caused by the lignin network. Since lignin has a hydrophobic network, the swelling of fibres and therefore the related decrease in the normalised DMT modulus was restricted. The green line in figure 11b shows an increase in the normalised adhesion/radius with an increasing RH for EP. Comparing these values to crystalline and amorphous cellulose, the increase is minor. It was interpreted that the more amorphous cellulose and its component parts are, the lower the increase in adhesion with increasing RH. Additionally, it was assumed that the hydrophobic lignin network reduced the attractive capillary forces between the tip and the EP fibre surface. Therefore, the adhesion increases in EP fibres were not as distinctive as those in LP fibres with amorphous and crystalline cellulose.

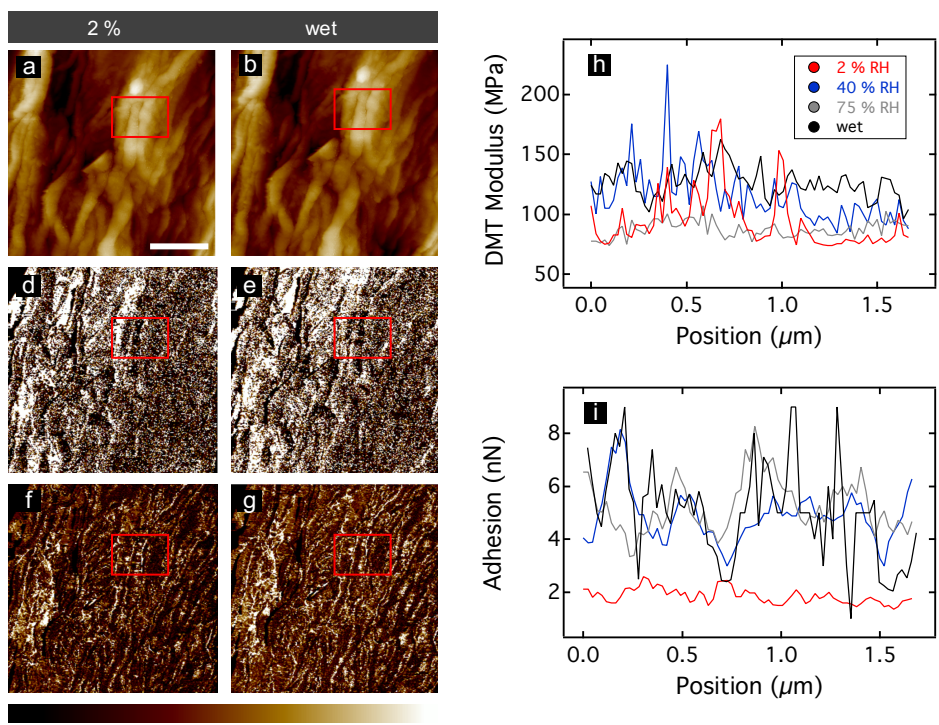

**Figure 12** AFM images of a PCEP fibre with a homogeneously coated area. a) Topography image at 2 % RH with the corresponding mechanical properties of d) DMT modulus and f) adhesion. b) The topography image under wet conditions with the associated mechanical properties of e) DMT modulus and g) adhesion. The scale bar represents 2 µm. The colour scale ranges from -550 – 550 nm for topography images, from 0 – 300 MPa for DMT modulus maps and from 0 – 10 nN for adhesion maps. h) and i) The cross sections of the DMT modulus and adhesion of the PCEP fibre.

Figure 12 shows the AFM images of a PCEP fibre. As seen in figure 12a, for 2 % RH, and in figure 12b, for the wet condition, there was no significant swelling of the microfibrils of the fibre in the topography images. This is interpreted as the first indicator of a hydrophobic polymer coating. By checking the mechanical properties such as the DMT modulus and adhesion in figure 12d,f for 2 % RH and figure 12e,g for the wet condition, they also revealed no significant changes when increasing the RH. Cross sections of the red framed areas in figure 12a-g are shown in figure 12h and i. These trends are pointed out in figure 11b, where the PCEP fibres are represented by the dashed green line. Even under wet conditions, the normalised adhesion/radius in PCEP fibres is less than that in plain EP fibres. The normalised DMT modulus of the PCEP fibres in 11a with the dashed green line shows almost no change within the error and therefore differs clearly from the EP fibres. This is interpreted as a strong indicator of the hydrophobicity of the polymer coating. The PCLP fibres are represented by dashed red lines in figure 11. Similar to the PCEP fibres, the PCLP fibres also exhibited almost no decrease in the normalised DMT modulus and no increase in the normalised adhesion/radius with increasing RH. The changes in the normalised DMT modulus due to the hydration recorded for the polymer-coated fibres were comparable to the values measured during the indentation experiments. Thus, it is concluded that the presence of a homogenous and stable hydrophobic polymer coating can be inferred from AFM measurements. Unmodified fibres exhibit large changes in mechanical properties, such as the DMT modulus and adhesion, as well as the swelling of microfibrils in topographic images. Hydrophobic coated paper fibres show small changes in mechanical properties and almost no swelling in topographical images when the RH was increased.

# 4 Conclusion

The terpolymer P(S-co-MABP-co-PyMA) was used as a water-repellent coating on cotton linters, eucalyptus fleeces and individual fibres. The polymer coating could be identified in scanning electron and fluorescence microscopy images. The CA revealed the hydrophobic character of the polymer coating at the macroscopic level. Fluorescence microscopy, however, showed an inhomogeneous distribution of the coating on the surface at the microscopic level. Raman spectroscopy revealed that the amount of water that was absorbed from a drop differed between the coated and uncoated samples. AFM experiments revealed mechanical differences between single LP or EP fibres and polymer-coated fibres in various states of hydration. On uncoated LP and EP fibres, an increased AFM indentation could be observed in the wet state, and the fibres became significantly softer. Compared with untreated fibres, fibres with a hydrophobic polymer coating showed a higher stiffness and less softening due to hydration.

The nanomechanical mapping also allowed for further insights into the local mechanics of coated and uncoated fibres. For uncoated LP fibres, crystalline and amorphous regions could be identified in the topography images, especially in the mechanical property maps. When increasing the RH, the topography images exhibited swollen microfibrils, and the microfibrils changed in size and shape. The swelling and therefore the softening of the fibre apparently started in the non-ordered regions. If both fibre types were coated with the hydrophobic polymer, no swelling of the microfibrils could be found in the topography images, and no significant changes in the mechanical properties occurred. In conclusion, Raman spectroscopy and AFM measurements revealed a softening of unmodified fibres in a humid environment due to an increased number of water molecules inside the cellulose network. The hydrophobization of the fibres with P(S-co-MABP-co-PyMA) effectively prevented swelling and mechanical weakening of the fibres. However, for the functionality of the coating, a uniform distribution of the polymer on the surface seems to be crucial.